\documentclass[10pt, conference]{IEEEtran}
\IEEEoverridecommandlockouts

\usepackage{subfigure}

\usepackage[utf8]{inputenc}
\usepackage{amsmath,amssymb}

\usepackage{graphicx}
\usepackage{subfigure}
\usepackage{color}
\usepackage{array}
\usepackage{url}

\usepackage{algorithm}
\usepackage{algorithmic}

\usepackage[colorinlistoftodos,prependcaption,textsize=small]{todonotes}
\usepackage{blindtext}
\presetkeys%
    {todonotes}%
    {inline,backgroundcolor=yellow}{}

\begin{document}

\title{Modeling the Internet of Things:\\a simulation perspective\footnotemark}

\author{\IEEEauthorblockN{Gabriele D'Angelo, Stefano Ferretti, Vittorio Ghini}
\IEEEauthorblockA{Department of Computer Science and Engineering, University of Bologna\\
Bologna, Italy\\
\{g.dangelo, s.ferretti, vittorio.ghini\}@unibo.it}
}


\maketitle

\footnotetext{
The publisher version of this paper is available at \url{https://doi.org/10.1109/HPCS.2017.13}.
\textbf{{\color{red}Please cite this paper as: ``Gabriele D'Angelo, Stefano Ferretti, Vittorio Ghini. Modeling the Internet of Things: a simulation perspective. Proceedings of the IEEE 2017 International Conference on High Performance Computing and Simulation (HPCS 2017)''.}}}

\begin{abstract}
This paper deals with the problem of properly simulating the Internet of Things (IoT).
Simulating an IoT allows evaluating strategies that can be employed to deploy smart services over different kinds of territories.
However, the heterogeneity of scenarios seriously complicates this task.
This imposes the use of sophisticated modeling and simulation techniques.
We discuss novel approaches for the provision of scalable simulation scenarios, that enable the real-time execution of massively populated IoT environments.
Attention is given to novel hybrid and multi-level simulation techniques that, when combined with agent-based, adaptive Parallel and Distributed Simulation (PADS) approaches, can provide means to perform highly detailed simulations on demand. 
To support this claim, we detail a use case concerned with the simulation of vehicular transportation systems.
\end{abstract}


\section{Introduction}
\label{sec:intro}

The Internet of Things (IoT) is firmly established today. We are surrounded by a multitude of sensors, devices, people equipped with mobile terminals, all somehow connected to the Internet, and the number of these things increases at a fast pace. 
Such a growing amount of devices can be employed as sources of sensed information, computation units, means for communication. It is thus important to devise strategies to let them interconnect \cite{Atzori:2010}.
These solutions must take into account that things may have very specific characteristics both in terms of hardware (in many cases, these devices are equipped with a very little amount of memory and computational power), software (specific OSes) and management (little or no administration utilities, few system updates). 

The wide spectrum of possible uses of things makes simulation a central tool for the real deployment of smart services. 
The complex networks obtained by the interaction of IoT devices are hard to design and to manage. 
IoT simulation is necessary for both quantitative and qualitative aspects. To name a few issues: capacity planning, ``what-if'' simulation and analysis, proactive management and support for many specific security-related evaluations. 
The problem is that modeling a general IoT environment can be quite difficult \cite{hpcs16,paper+kirsche-13:iot-simulation}. 
There is wide a number of different aspects to take into consideration. Among them,
scalability is a main one. Traditional approaches (that are single CPU-based) are often unable to scale to the number of nodes (and level of detail) required by the IoT. 

This paper introduces the main aspects of the simulation of IoT, discussing a new combination of techniques to enhance scalability and to permit the real-time execution of massively populated IoT environments (e.g., large-scale smart cities). For example: parallel and distributed simulation (PADS), adaptive computational and communication load-balancing, self-clustering. In particular, attention is given to the hybrid and multi-level modeling and simulation techniques. 
In few words, a hybrid simulation is a simulation where multiple simulation models are glued together~\cite{magne2000towards}. Each simulator has a specific task, and these simulators are somehow orchestrated by some simulation coordinator. We also refer to multi-level simulation when these simulation tools work at a different level of detail \cite{hpcs16,gda-simpat-iot}.
These solutions allow creating multiple, interacting instances of different simulations, that are specifically designed to focus on particular aspects in a reduced portion of the simulated area, or on a reduced subset of simulated entities.

To demonstrate the validity of the proposed approach, we analyze an application scenario related to the simulation of vehicular transportation systems.
The classic ways to analyze vehicular networks relate to two antithetical approaches. One is an abstract simulation of vehicles as moving entities, where these entities are simple agents moving in a constrained simulated space. Thus, we label the agent as a vehicle and force it to move in given paths, representing streets. But from a simulation point of view, the same model might represent ants moving over tree branches. Thus, no details related to vehicular systems are considered.
The other approach is to design a detailed simulation, embodying all sophisticated aspects of the technologies inside a modern vehicle, i.e., motor, pollution-related aspects, networking technologies to connect to the Internet, etc. Due to all these details to consider, the typical simulation can only be composed of few entities, due to scalability issues concerned with the computational costs for mimicking a vehicle.

As a matter of fact, there are situations in which the use of the first approach can introduce errors due to the oversimplification, while in other situations the latter approach does not scale due the presence of unnecessary details. Hybrid and multi-level approaches can solve the problem, since they allow designing and configuring smart services in large scale vehicular transportation systems over wide area networks. Detailed simulations can be triggered nevertheless. The advantage is that the detailed (and thus, more costly) simulation can be performed only when needed, in a limited simulated area, only for the needed time interval of the simulation.

The remainder of this paper is organized as follows. Section \ref{sec:background} describes the background about simulation techniques, useful to introduce aspects related to the simulation of IoT. In Section \ref{sec:stateoftheart} the state of the art related to IoT simulation is discussed. An approach based on adaptive parallel/distributed simulation and multi-level simulation is introduced in Section \ref{sec:multilevelsimulation}. In Section \ref{sec:casestudy}, this approach is applied to a case study on intelligent transportation systems. Finally, Section \ref{sec:conc} provides some concluding remarks.

\section{Background}
\label{sec:background}

\subsection{Simulation and Discrete Event Simulation}
In a computer simulation, a process models the behavior of some other system over time~\cite{FUJ00}. The system to be simulated can be already existing or yet to be built. In most cases, the simulation tools are used to support the design and implementation of complex systems. In fact, most modern systems are so complex that their study (i.e.~dimensioning and tuning) can be done only by using simulations.

Among the many simulation paradigms that have been proposed, Discrete Event Simulation (DES)~\cite{Law:1999:SMA:554952} is very popular for its combination of expressiveness and usability. More in detail, the main component of a DES is the simulation model, that is implemented by a set of state variables. These variables represent the simulated system in a given moment of its evolution. Such evolution of the simulated model is obtained processing an ordered sequence of events. Each event represents a change in the simulated model state and it is tagged (i.e.~timestamped) with a specific simulated time. In other words, the simulation evolution is obtained computing an ordered sequence of events that need to be created, stored in a specific data structure and then processed using the appropriate processing handlers. As an example, the simulation of a vehicular ad hoc network is made by events that represent the different car positions during the simulation and the transmission of wireless packets.

The implementation of DES is usually made by i) a set of state variables, that are used to represent the state of the modeled system, ii) an ordered event list, that are the pending events waiting to be processed to evolve the simulation, iii) a global clock, that represent the current time in the simulated system~\cite{Law:1999:SMA:554952}.

\subsection{Sequential DES}
When a single CPU core (also called a Physical Execution Unit, PEU) manages the whole simulated model and its evolution, the simulation is defined as sequential (i.e.~monolithic). This means that the PEU is in charge of processing all the events in the correct timestamp order, originating new events and updating the pending event list. The processing of events in the correct order is necessary in order to avoid causality errors in the simulation. The main advantage of the sequential approach is that it is easy to implement and to debug but there are also some drawbacks. For example, large systems require a huge number of events to be stored and processed. Since all of them are sequentially executed by a single PEU, then the scalability of this approach is limited and the amount of time required by the simulation runs is often excessive~\cite{1668384}. 

\subsection{Parallel DES and PADS}
An alternative approach to sequential DES is called Parallel Discrete Event Simulation (PDES). In this case, a set of interconnected PEUs (e.g.~CPU cores or hosts) is in charge of the simulation execution~\cite{Fujimoto:1989:PDE:76738.76741}. The simulation model is partitioned among different PEUs and each of PEU is in charge of representing and executing only a part of the whole simulation model. 
More in detail, each PEU implements a local pending events list but some events need to be delivered to other PEUs using a message passing approach. The partitioning of the simulation model can increase the simulator scalability (thanks to the parallelization of some tasks) but the set of PEUs needs to be properly synchronized to prevent causality errors. A PDES can be faster than the corresponding sequential DES in simulating the same model, but this happens at the cost of a more complex implementation and management of the simulator.

\begin{figure}[ht]
\centering
\includegraphics[width=\linewidth]{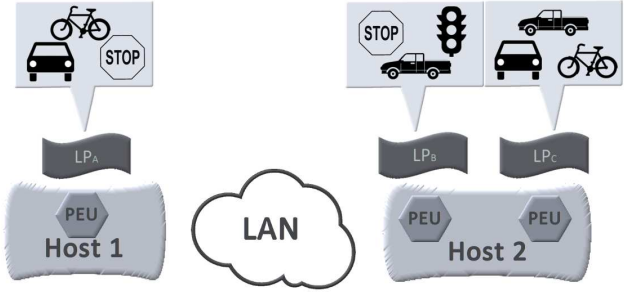}
\caption{Model partitioning in the simulation of a smart city.}
\label{fig:pads-model}
\end{figure}

The definition of Parallel and Distributed Simulation (PADS) provided in \cite{perumalla2007} is quite simple: ``a simulation that is run on more than one processor''. Reduced time of the simulation runs (with respect to the sequential approach), model and simulator scalability, interoperability of simulators and composability of simulation models are among the many advantages of PADS with respect to sequential simulators~\cite{FUJ00}. In the PADS terminology, each PEU implements a model component (called Logical Process, LP) that is a part of the whole simulation~\cite{gda-simpat-2017}. In other words, a PADS is made by the LPs and their interactions (see Figure~\ref{fig:pads-model}). In fact, each LP manages the evolution of a part of the simulated model and communicates with other LPs for the necessary synchronization and data distribution tasks~\cite{FUJ00}.

The main characteristics of PADS (with respect to sequential DES) is the lack of a global model state. In other words, in the PADS execution architecture the single node in which the whole simulation model is stored (and managed) is missing. In fact, in this case, the simulation evolution is obtained only through the coordinated computing and communication of nodes arranged in a parallel/distributed architecture.

It is not always easy to define what is the difference between a parallel and a distributed simulation. In this paper, for the sake of simplicity, we assume that when the PEUs are interconnected by shared memory then it is a parallel simulation and when the PEUs are connected by LAN (or Internet) then it is distributed. Nowadays, most execution architectures are a mix of parallel and distributed components~\cite{gda-simpat-2017}.

Clearly, the characteristics of the network that interconnects the PEUs have a strong effect on the PADS performance. For example, the latency and bandwidth constraints in LAN-based communications (or Internet) slow down the simulation execution with respect to the low latency and high bandwidth that can be found in shared-memory multi-processor.

To summarize, the main characteristics of PADS are:
\begin{itemize}
	\item the simulated model has to be partitioned in a set of LPs~\cite{bagrodia98}. The \textbf{partitioning} is a multi-objective optimization problem in a dynamic system in which a part of the information is unknown a priori. More in detail, the partitioning of the simulated components in the parallel/distributed architecture must be done minimizing the amount of network communication among the LPs while load-balancing (i.e.~computation and communication) the execution architecture;
		\item \textbf{correctness}: the PADS results are correct if and only if they are the same as those of the corresponding sequential simulator. Since each PEU has different hardware characteristics and executes a specific part of the simulation model, it is necessary to coordinate the PEUs using an appropriate \textbf{synchronization} algorithm;
	\item since the simulated model is partitioned in different LPs, some of the updates (i.e.~events) that are generated in a specific LP can be of interest for the components that are allocated in other LPs. A simplistic solution would be to broadcast all the updates. This behavior would introduce a massive communication overhead and therefore it is not acceptable. \textbf{Data distribution} is about the efficient delivery of state updates. This can be achieved in many ways, among them publish-subscribe mechanisms~\cite{Jun:2002:ESM:564062.564074}.
\end{itemize}

The implementation of PDES using a PADS approach is obtained encapsulating the events in timestamped messages. To obtain a correct execution of the PDES, the delivery and processing of such messages must be done accordingly to the causality constraint: ``two events are in causal order if one of them can have some consequences on the other''~\cite{Lamport1978}. In other words, a correct PDES execution is obtained when the causal order of events is not violated. In the case of a DES, with a sequential execution (and a single pending event list), it is easy to avoid causality violations. On the other hand, in complex parallel and distributed architectures, there are PEUs with different execution speeds, network delays, model and execution architecture imbalances. In practice, in a PADS, a synchronization algorithm is needed for the correct handling of the simulation execution. In the years, many different synchronization algorithm have been proposed; in the following we summarize the main approaches that can be followed:

\begin{itemize}
	\item \emph{time-stepped}: the simulated time is represented as a sequence of fixed-size timesteps. The timestamp associated to events is the timestep in which they have to be processed. In other words, the simulation model is updated at every timestep. A consequence of this approach (that is also a relevant limitation) is that the timestep size is the lower bound to the interaction between model components. In other words, each interaction between two separate model components is not instantaneous but requires a certain amount of time to be delivered. The implementation of the time-stepped synchronization mechanism can be centralized or distributed. In the distributed one, when a LP completes the processing of the current timestep then it broadcasts an End-Of-Step (EOS) message to all other LPs. Each LP waits to collect all the EOS messages (from other LPs) before jumping to the following timestep~\cite{1261535};
	\item \emph{conservative}: the main assumption of this approach is that causality violations must be prevented. This means that each event must be analyzed before its processing. If the event is ``safe'' (in terms of causality violations) then it can be processed. Otherwise, the LP must stop the processing of this event and switch to the evaluation of other events. In the case where all the events are unsafe then the LP must wait for more information on the safeness of waiting events. Many synchronization algorithms implement this approach, among them the Chandy-Misra-Briant~\cite{misra86} and its variants are quite popular;
	\item \emph{optimistic}: this approach is about processing the events in receiving order without assessing their causality constraints. This means that causality violations may happen and so the mechanism must be able to find the violations and fix them. In the Time Warp algorithm~\cite{timewarp} this is done implementing a roll-back mechanism of the LP state variables and (if necessary) propagating the roll-back to all the LPs that have been affected by causality violation. Such cascade of roll-backs brings the state of the whole PADS back to the most recent one that is free from causality violations. This restored state is then used by the LPs as the new starting point for the processing of events.
\end{itemize}

\subsection{Adaptive PADS}
As described above, the model partitioning is one of the main problems of PADS. Most of the approaches that can be found in the literature rely on a static partitioning (and clustering) of the simulated model components in the available LPs. To overcome the limitations of a static approach, in~\cite{gda-simpat-2017} we proposed a dynamic mechanism that is based on the self-clustering of model components. More in detail, the model is partitioned in a large set of small-sized model components that are called Simulated Entities (SEs). In this case, the model evolution happens through the interactions among SEs. It is easy to see that this approach is very similar to a multi-agent system. In an adaptive PADS, the LPs are containers of SEs and then it is possible to manage the migration of SEs from one LP to another. The adaptive reallocation of SEs allows the reduction of the communication overhead in the parallel/distributed architecture (e.g.~clustering the simulated components that interact with high frequency in the same LP) while load-balancing the execution architecture. In many simulated models (and execution architectures), this approach can lead to a speedup of the execution runs and a better scalability.

The GAIA/ART\`IS simulator~\cite{pads} implements the adaptive PADS mechanism described above on top of the time-stepped synchronization. In our previous work~\cite{gda-simpat-iot} we have demonstrated that this approach can be integrated in a multi-level simulation. In the following of this paper we will see that even a hybrid simulation approach can be used in combination with the adaptive PADS.

\subsection{Hybrid Modeling and Simulation}
The scientific literature is missing a clear and cohesive definition of hybrid simulation~\cite{Eldabi:2016:HSH:3042094.3042274}. Despite this, the mixing of analytic and simulation models is not new~\cite{10.2307/170837}. In fact, in the past this approach has been already implemented in a few simulators~\cite{Mosterman1999}. In this paper, we consider hybrid models all the solutions in which there is interoperability of simulation models that follow different simulation approaches, For example, linking discrete event simulation (DES) with either system dynamics (SD) or agent based (ABS)~\cite{Eldabi:2016:HSH:3042094.3042274} but also analytical models (e.g.~continuous simulation).

\section{State of the Art}
\label{sec:stateoftheart}

\subsection{Simulation of the Internet of Things}

The design, setup and tuning of large scale IoT systems requires the availability of either testbeds or simulation tools~\cite{7879128}. In many cases, the cost and complexity of such testbeds is so high that simulation is the only option. The simulator used for the performance evaluation must be able to deal with massively populated IoT environments and a high level of detail in the interactions among the simulated components. Both these aspects are fundamental for the scalability of the simulation tools and must be properly considered when reviewing the state of the art.

In~\cite{6069710}, the authors consider both the aspects discussed above, in fact they identify the main requirements for the experimental facilities needed for the design and evaluation of future IoT deployments. A discussion of the main drawbacks of simulation-based approaches is followed by a survey of existing IoT testbeds. The authors show that an approach based on the federation of testbeds is feasible but with drawbacks. It is worth noting that some testbeds are able to support an approach that integrates simulation components, a solution is often referred to co-simulation. One of the conclusions of this paper is that the existing network simulators are unable to support the scale and the level of detail required by future IoT systems.

To improve the simulator scalability, SimIoT~\cite{6844677} uses a cloud environment for the execution of back-end operations. In \cite{6844677}, an use case is described, based on a health monitoring system for emergency situations. The performance evaluation considers 160 identical jobs that have been submitted by 16 IoT devices. More complex setups are needed to assess the scalability of the proposed approach.

The problem of the large number of devices that must be considered in IoT deployments is discussed in \cite{6664581}. In this paper, the authors firstly overview the available large-scale simulators and emulators. Secondly, they propose MAMMotH, a software architecture based on emulation. This approach is promising, but the development of MAMMotH seems to have stopped in 2013.

The integration of a general-purpose discrete event simulation (e.g.~DEUS) with domain specific simulators (e.g.~Cooja and ns-3) is pursued in \cite{Brambilla:2014:SPL:2694768.2694780} for assessing large-scale IoT setups in urban environments. In this case, the authors consider 6 scenarios of medium complexity (i.e.~$200.000$ sensors, $400$ hubs and $25.000$ vehicles). The performance evaluation of the integrated simulator shows a good scalability even if DEUS is Java-based and with a monolithic architecture.

The integration of different simulators is part of the approach proposed in \cite{paper+kirsche-13:iot-simulation}, in which Cooja-based simulations (i.e.~system level) are binded with a domain specific network simulator (i.e.~OMNeT++) to obtain a hybrid simulation environment.

In \cite{Wehner2017}, the OMNeT++ simulation framework is used to model an IoT network infrastructure composed of sensors, actuators, and even processors. This approach, that is again based on a domain specific simulator, permits the simulation of components (that are not yet available) and even the presence of hardware in the loop.

\cite{wrro111418} proposes what the authors call an Internet of Simulation (IoS) that is a set of interconnected simulations in which all the models and simulations are exposed to the Internet and can be accessed on an ``as-a-service'' basis (i.e.``simulation as a service'').
 
An interesting solution is proposed in \cite{Brumbulli2016}. In this case, the SDL language is used to model the IoT scenario. In the following, an automatic code generation is in charge of translating the SDL description in a ns-3 simulator model. The clear limitation of this approach is the scalability of ns-3.

In \cite{fortino2016simulation} is discussed the integration between an agent-based methodology and a domain specific simulator (i.e.~OMNeT++). Differently from the approach described in this paper, the agent-based methodology is used for the modeling while OMNeT++ implements the simulations.

\subsection{Internet of Things and Smart-Cities}

Concerning the use of IoT to build efficient services for making ``smarter'' territories, from a simulation point of view
there are many requirements that the simulation tool must provide. 
There are several parameters involved that should be considered and possibly varied in order to perform a what-if analysis. These parameters might force the IoT set-up and deployment. For instance, assume that you need to create a Wireless Sensor Network (WSN) of interacting in a IoT, employed to build a smart service. In this case, a parameter might be the wireless transmission ranges of communicating devices. This parameter is influenced by the geographical location where the WSN has to be deployed, and in turn the transmission range influences the amount of sensors to be deployed in the area.

Above all, the main issue is scalability, both in terms of amount of modeled entities and granularity of events. An IoT will be composed by thousands of interconnected devices. Many of them will be mobile and each with very specific behavior and technical characteristics~\cite{hpcs16,smartshires}. Then, in certain scenarios, (almost) real-time simulations are required. This is the case when proactive approaches are utilized to perform ``what-if analyses''.

Hybrid and multi-level simulation enable the simulation of smart territories based on the use of the IoT. In fact, running a complex, massively populated model at the highest level of detail is unfeasible. A more profitable solution is to organize the simulation as an orchestration of multiple simulators. Each single simulator focuses on a specific level of detail, with specific characteristics of the domain to be simulated (e.g.~mobility models, wireless/wired communications and so on). 

Agent-based simulation is the typical tool employed to mimic urban systems, smart cities and transportation systems~\cite{Karnouskos}. Agent-based simulation, together with land-use transport interaction model and cellular automata are applicable in planning support systems. Different time scales can be modeled; for instance, one can perform short-term simulations to model diurnal patterns in cities, while longer term models can be exploited for strategic planning purposes. MASON~\cite{Luke:2005} and SUMO~\cite{SUMO2012} are examples of simulation tools for the simulation of moving (e.g.~mobile users, vehicles) or static entities. These tools have been successfully exploited to study intelligent traffic control systems~\cite{bauza,kerekes,Wegener:2008,e16052384}, mobile applications that resort to crowdsensed data~\cite{PrandiFMS15} and so on. While the implementation of models is quite simple for a generic programmer, these approaches do not allow creating massive scenarios, with many interconnected things.

CupCarbon is a multi-agent and discrete event simulator, thought to simulate smart-cities and IoT WSN~\cite{Mehdi:2014}. In particular, its purpose is to enable distributed algorithms validation. This tool employs OpenStreetMap to simulate sensors deployment on a map. The main goal of this tool is to help trainers to explain the basic concepts and how sensor networks work and it can help scientists to test their wireless topologies, protocols, etc. The main problem of scalability remains.

We conclude this section mentioning simulators that are more prone to image and 3D based representations of smart cities. Examples are CanVis, Second Life, Suicidator City Generator, Blended Cities. In particular, it is worth mentioning UrbanSim, that provides tools for examining the interplay between land use, transportation, and policy in urban areas~\cite{urbansim}. It is intended for use by Metropolitan Planning Organizations and others needing to interface existing travel models with new land use forecasting and analysis capabilities. UrbanSim does not focus on scenario development, as most of these tools do, but rather on understanding the consequences of certain scenarios on urban communities. However, such a kind of tools do not usually cope with issues concerned with wireless communications and pervasive computing, which are the keywords related to the IoT world.

\section{Multi-level Hybrid Simulation}
\label{sec:multilevelsimulation}

The fine grained simulation models needed for the accurate assessment of IoT have scalability problems in presence of a large number of nodes, as typical in IoT systems. In other words, the time required by a monolithic simulator to obtain statistically correct results is excessive. Furthermore, we need simulation tools that can be used for the real-time assessment and ``what-if analysis'' of complex IoT setups.

An approach based on PADS can enhance the simulator scalability but with some significant limitations, in fact massively populated IoT would still be difficult to handle. The common solution to the simulators scalability problem is to reduce the level of detail in the simulation model. In our view, this solution might turn out to be very dangerous, in terms of simulation outcomes. In fact, it often leads to misleading (or wrong) results. An alternative solution, that we have proposed in the past~\cite{gda-simpat-iot}, is based on multi-level modeling and simulation~\cite{Ghosh:1986:CDM:319541.319559}. In the approach that we propose, multiple simulation models (and simulators) are the components of a new combined simulator~\cite{magne2000towards}. More in detail, each component is able to handle a specific task and to work at a different level of detail (i.e.~multi-level simulator). In addition, each simulator can follow a different simulation paradigm (i.e.~hybrid simulator).

As an example, a ``high level'' adaptive PADS simulator (i.e.~GAIA/ART\`IS) can be used to coordinate the execution of some domain specific simulators (e.g.~OMNeT++~\cite{omnet}, ns-3~\cite{ns3}, SUMO~\cite{sumo}). The ``high level'' works at a coarse grained level of detail, while the ``low level'' simulators are used for the fine  grained of some specific parts of the simulated system. The switch between ``high level'' and ``low level'' can be automatic (e.g.~based on specific locations in the simulated area) or triggered by the simulation modeler (e.g.~for the detailed analysis of specific behaviors observed during the simulation). For example, the presence of hotspots of wireless devices in a simulated area can cause network capacity and congestions problems that need to be analyzed with specific simulation tools.

\begin{figure*}[ht]
\centering
\includegraphics[width=.7\linewidth]{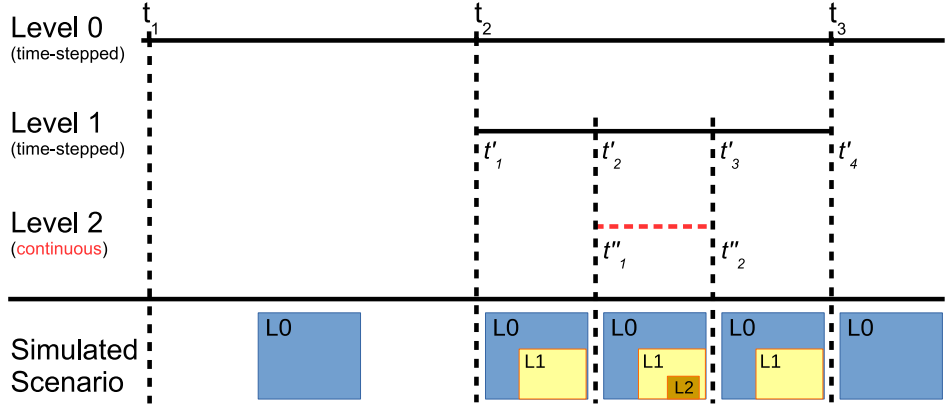}
\caption{Multi-level hybrid simulation scheme, the level 0 and 1 simulators are time-stepped while the level 2 is continuous. Each simulator works at a different level of abstraction.}
\label{fig:multilevel-simulation}
\end{figure*}  
  
The main issues to cope with, when dealing with hybrid/multi-level approaches, are the interoperability among the simulators and the design of the inter-model interactions. In fact, such interactions impose synchronization and runtime communication of state exchanges among model components.

Figure~\ref{fig:multilevel-simulation} shows an example of a hybrid/multi-level simulation scenario. At the simulation bootstrap, the whole simulation is performed at level 0 (hence, with minimal details). This means that the high level simulator is in charge of all the model components and their interactions. As said before, the level 0 simulator implements time-stepped synchronization~\cite{gda-simpat-2014}. When a specific portion of the simulated area needs to be simulated with a higher level of detail (i.e.~at timestep $t_2$), then another simulation level is triggered (only for that specific simulation area). This means that the state of a specific group of simulation components needs to be transfered from level 0 to level 1. The result is that a part of the simulated area is still simulated at level 0 while a specific zone is implemented at level 1. Following this approach, different simulation areas can be simulated at different levels of detail, concurrently. It is clear that different simulation tools (following different modeling approaches) can be used for the different areas. If needed, then a sub-portion of level 1 simulation might be simulated at level 2 (that is an even higher level of detail). In this specific case, the simulator at level 2 does not follow a time-stepped approach; that is, it implements a continuous simulation.

To simplify, we will detail this description considering only the first two levels (i.e.~level 0 and 1). All the model components managed by the level 0 simulator are synchronized with $t$-sized timesteps and all level 1 components with $t'$-sized timesteps. Timestep $t_2$ (i.e.~$t'_1$ at level 1) is when there is the switch of some model components from the coarse grained simulator to the finer one. Going on with the simulation execution, the level 0 components will jump from $t_2$ to $t_3$ while the level 1 components will update their state $t'_2$, $t'_3$ and $t'_4$. It is worth noticing that $t'_4$ at level 1 is the same of $t_3$ at level 0. At $t'_4$, the execution of the level 1 simulator is terminated and all its simulation components must be migrated back to the higher layer (i.e.~level 0). All that procedure needs to be arranged under the constraints of the time-stepped synchronization mechanism. This means that all the interactions among level 0 components must happen only at coarse grained timestep while the interactions at level 1 happen at fine grained timestep. Finally, the interaction between components managed at different levels need to be arranged at the coarse grained timesteps, that is when there is a match between the timesteps at the different levels.

During a hybrid/multi-level simulation, the total number of simulated entities might not change; what changes is the level of detail used to perform the analysis. This clearly increases the scalability of the whole simulation system, since in-depth simulations are performed only when needed and for a subset of entities. It should also be clear that higher level simulations might introduce some errors, due to the lack of detail. Thus, the trade-off here becomes when (in the simulated time) and where (in the simulated area) triggering more detailed simulations (higher simulation costs), rather than keeping a simplified simulation model (larger approximation errors). As in every simulation, appropriate verification and validation techniques need to be used.

\section{A Case Study with Intelligent Transportation Systems}
\label{sec:casestudy}

An important use case for the IoT lies in the smart cities domain and relates to transportation systems.
Numerous examples exist of startups, services and technologies being developed. Just to mention a few, 
BestMile is a cloud platform to manage autonomous vehicle fleets. 

Kiunsys is developing solutions to deal with all aspects of parking, ranging from analytics software to sensors management. Based on this solution, the city of La Spezia (Italy) has deployed more than 1000 parking spot sensors to communicate free parking spots in real time. 

Hi-Park is another example of parking application.

Anagog has recently built a platform enabling mobile applications developers to collect and analyze in real-time raw signals from multiple smart-phone sensors, in order to determine and predict the user mobility status.
Thus, for instance, data coming from a smart-phone can be used to determine whether a user is driving a car. Based on this information, it is possible to deliver assistive services and information, or stop those that could distract drivers.

The Array of Things is a more general urban sensing project, which has some strong implications on transportation systems. It builds a network of interactive, modular sensor boxes to collect real-time data on the city environment, infrastructure and activity for research and public use. It has been installed in the city of Chicago (USA). The goal is to use this technology as a ``fitness tracker'' for the city, measuring factors that impact on livability in Chicago such as climate, air quality, noise and to use it to regulate traffic.

The modeling and simulation of an urban (or rural) scenario, equipped with a large amount of sensors, devices and mobile nodes that produce data to be used in intelligent transportation systems, requires taking into account several issues. These range from data gathering and distribution, communication and interaction among vehicles, up to path planning strategies and pollution issues.
It is thus evident that the use of a multilevel/hybrid simulation approach may introduce serious advantages.
In fact, assuming the need to consider traffic conditions of a geographical area, it is necessary to take into account the street map, identify the critical points and understand if it is possible to tune or modify the traffic circulation of the area, also considering polluting emissions.
While it might seem prohibitive to take all such issues in a single simulator, the composition and properly tuned interaction of different simulators can solve the problem.
To devise a solution, we will use a top-down approach, starting from a higher level of abstraction that models the whole general area, and then describing more detailed solutions for specific problems, to be simulated in smaller regions of the simulated area.

\subsection{Level 0: Modeling the urban area}

The urban street area can be viewed as a complex network. This, through a mathematical analysis, allows the analysis of the whole road network and the identification of the network characteristics, the shortest-path routes, the diameter of the net, the critical points (e.g., those intersection points that have high centrality measures). Moreover it allows to find the presence of cul-de-sacs, calculate statistics like intersection density, average node connectivity, etc.

A tool that can be used to obtain this is OSMnx~\cite{Boeing16}. 
OSMnx is a Python package that is able to retrieve administrative boundary shapes and street networks from OpenStreetMap, and export it using a typical representation employed by complex network software tools, such as NetworkX or Gephi. 

\begin{figure*}[ht]
\centering
\includegraphics[width=\linewidth]{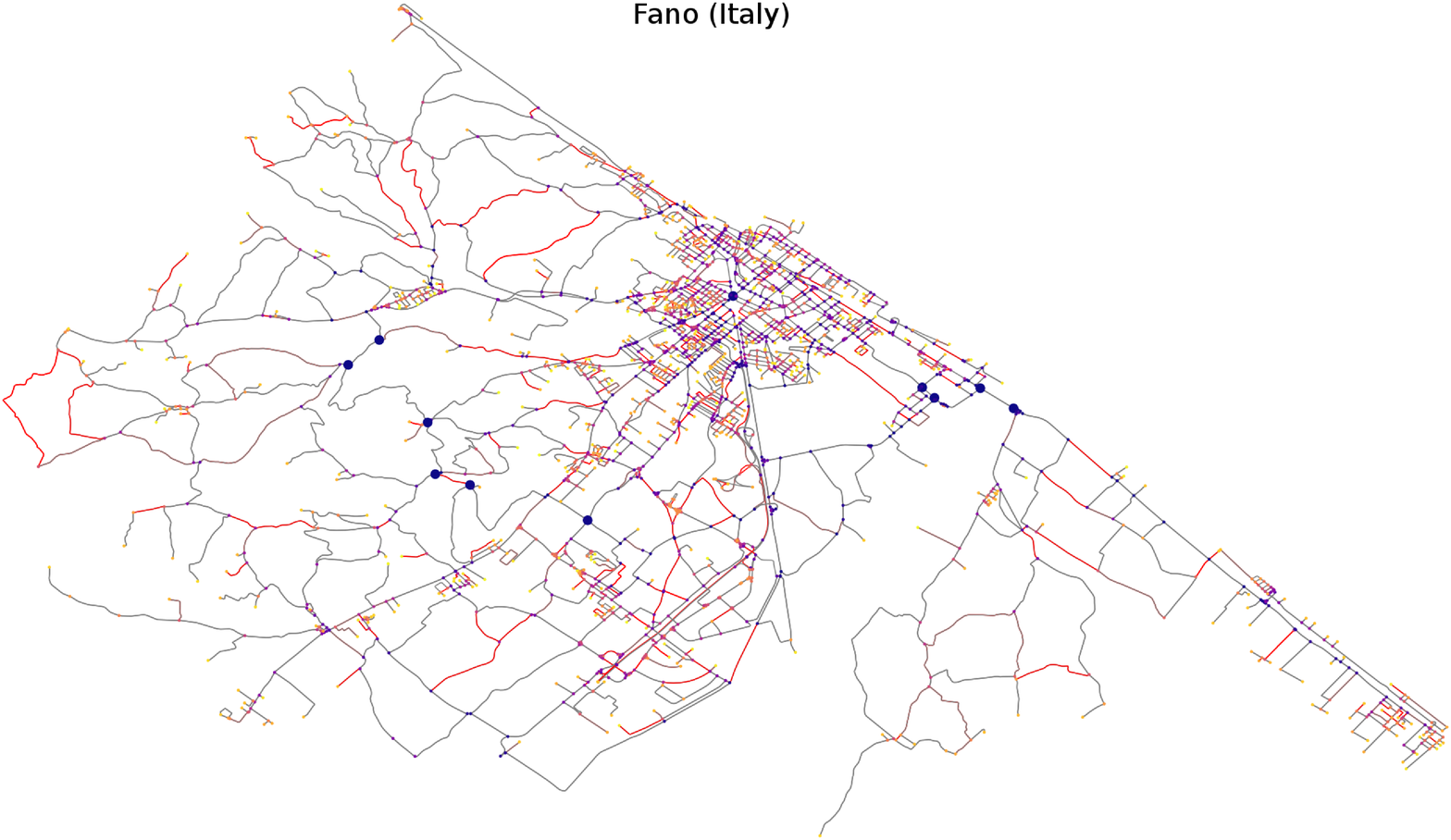}
\caption{Street Map of Fano (Italy). Gray lines represent two-way streets, red lines one-way streets. Crossroads are colored based on their betweenness centrality level, the darker the higher. Nodes bigger in size represent the crossroads with highest betweenness (i.e., critical points).}
\label{fig:fano}
\end{figure*}

Figure \ref{fig:fano} shows a visual representation of the street map of the city of Fano (Italy). In the map, gray lines represent two-way streets, while red lines are one-way streets. 
Nodes are crossroads, traffic circles, semaphore crossings; these are colored based on their betweenness centrality level (the darker the higher value). 
Nodes which are bigger in size represent crossroads (or traffic circles) with highest betweenness. 

Betweenness is 
a measure of centrality based on shortest paths. In particular, given a node $n$, the betweenness centrality of $n$ measures the amount of shortest paths among all pairs of network nodes $(x,y)$, passing through $n$, with respect to all shorted paths from $x$ to $y$.
In other words, betweenness centrality represents the degree of which nodes stand between each other.
Hence, it is a measure that helps identifying the critical points in a network.

Based on this preliminary analysis, the modeling tool identifies critical points and triggers more specific traffic related simulations.

\subsection{Level 1: Simulation of the urban area}

The step described above is useful to properly understand which are the places that the hybrid simulation tool has to monitor in detail. In case of a large urban area to monitor, the focus on specific critical points allows reducing the computational costs to perform the whole analysis.
In order to perform a simulation of the whole urban area, several simulators can be exploited. Agent-based simulators can be of real help in this case~\cite{hpcs16}.
In particular, one can employ the well known SUMO as a tool for the road traffic simulation \cite{sumo}, or some properly network assessment tools built over an agent-based simulation with PADS capabilities~\cite{gda-jpdc-2017}.

This level is thought to control and validate the road network and to perform what-if analyses by varying the amount of vehicles, assessing traffic circulation when introducing barriers, removing roads or adding novel ones.
The simulator at this level will be in charge of coordinating the execution of a number of lower level simulators, to study the goodness of transportation solutions. 
Regarding the specific traffic analysis, it might be possible to trigger a more detailed simulation, still based on a SUMO-like solution, for instance.
Moreover, it is possible to trigger simulations, on smaller portions of the area, assessing tailpipe emissions and pollution problems in general.
Communication issues in infrastructured and infrastructure-less wireless networks can be simulated through specific simulation tools, e.g., Omnet++ based.
Finally, in order to properly simulate smart cities' services, it might be possible to resort to some cloud simulator (see Section~\ref{sec:cloudsim}).

\subsection{Level 2a: Environmental simulator}

All the issues concerning the impact of the vehicular traffic on the general environment can be simulated by resorting to tools such as the ADvanced VehIcle SimulatOR (ADVISOR)~\cite{advisor}.
ADVISOR is a MATLAB/Simulink based simulation tool for the analysis of the performance and fuel economy of conventional (gasoline/diesel), electric, and hybrid vehicles. 
It allows interchanging a variety of components, vehicle configurations, and control strategies.
The goal of the simulator is to allow testing efficiency of automobiles, especially in terms of tailpipe emissions, fuel economy, acceleration and grade sustainability.
To this aim, the simulator works using a component-based approach, where components are typically modeled through a set of equations and quasi-steady approximations.
While the typical use of the tool is based on a graphical interface, it provides means to perform batch simulations. This eases the interaction with other components of a hybrid simulation software.

Using ADVISOR, it is possible to build a simulator that, based on the vehicles present on a given portion of the considered geographical area, measures the amount of emissions. These results would be passed to the higher level simulator.

\subsection{Level 2b: VANETs and Vehicular communications}

Vehicular communication networks might be based on some general networking infrastructure, or on some ad-hoc solution.
Tools such as VANET Omnet++ implement an intelligent transportation system and allow utilizing various types of vehicle communication, such as Vehicle-to-Vehicle (V2V) communication and Vehicle-to-Infrastructure (V2I) wireless communications. 
Such a tool allows generating a topology of vehicles equipped with one (or more) network interface card and using some communication protocols.

At this level, it is possible to study all problems concerned with the deployment of a networking infrastructure for supporting vehicular communications, as well as more sophisticated solutions. For instance, in case of intermittent connections, seamless communication strategies that employ multi-homing mechanisms might be tested~\cite{Ferretti2016390}.

\subsection{Level x: Simulating the cloud}
\label{sec:cloudsim}
As said above, most of the IoT architectures that will be deployed in the next years will be cloud-based. This means that a comprehensive modeling and simulation approach must consider the problems involved with the simulation of complex cloud services. In other words, it will not be acceptable to simulate only the low-level part of the IoT software architecture (e.g.~devices, sensors and communication networks). In fact, most of the performance (and scalability) of IoT services will depend on the cloud services that will provide essential coordination and communication services. For this reason, we are working to integrate simulators such as CloudSim~\cite{cloudsim} in our multi-layer/hybrid simulation architecture. In the following, this will permit to consider emerging testbeds such as edge and fog computing environments~\cite{DBLP:journals/corr/GuptaDGB16}.

\section{Conclusions}
\label{sec:conc}

This paper presented the main issues that arise in the simulation of the Internet of Things and in the deployment of smart services on smart territories. The two main issues are the need for scalability and high level of detail in the simulation.
However, these two requirements lead to technical solutions that are counterposed. In other words, you typically have to trade the high level of details for scalability.
We also provided an overview of the existing simulation techniques, reaching the conclusion that a good strategy relies on the use of adaptive, agent-based, Parallel and Distributed Simulation (PADS), coupled with multi-level and hybrid simulation approaches.

To clarify the advantages of a hybrid, multi-level simulation approach, we presented a use case related to intelligent transportation systems. 
In this case, wide geographical areas, with a multitude of simulation entities, can be simulated with agent-based PADS. However, when needed it is possible to trigger a more detailed, fine grained simulation, so as to consider aspects which could not be simulated otherwise. The interesting aspect of this approach is that the detailed (and more costly) simulation can be performed in a specific, limited simulated area, only for the needed time interval of the simulation.

\small{
\bibliographystyle{abbrv}
\bibliography{paper}  
}


\end{document}